# Principal component analysis for big data

Jianqing Fan,[*] Qiang Sun,[†] Wen-Xin Zhou[‡] and Ziwei Zhu[§]

**Abstract**

Big data is transforming our world, revolutionizing operations and analytics everywhere, from financial engineering to biomedical sciences. The complexity of big data often makes dimension reduction techniques necessary before conducting statistical inference. Principal component analysis, commonly referred to as PCA, has become an essential tool for multivariate data analysis and unsupervised dimension reduction, the goal of which is to find a lower dimensional subspace that captures most of the variation in the dataset. This article provides an overview of methodological and theoretical developments of PCA over the last decade, with focus on its applications to big data analytics. We first review the mathematical formulation of PCA and its theoretical development from the view point of perturbation analysis. We then briefly discuss the relationship between PCA and factor analysis as well as its applications to large covariance estimation and multiple testing. PCA also finds important applications in many modern machine learning problems, and we focus on community detection, ranking, mixture model and manifold learning in this paper.

[*]Department of Operations Research and Financial Engineering, Princeton University, Princeton, NJ 08544, USA. E-mail: jqfan@princeton.edu.

[†]Department of Statistical Sciences, University of Toronto, Toronto, ON M5S 3G3. E-mail: qsun.ustc@gmail.com.

[‡]Department of Mathematics, University of California, San Diego, La Jolla, CA 92093. E-mail: wez243@ucsd.edu.

[§]Department of Operations Research and Financial Engineering, Princeton University, Princeton, NJ 08544, USA. E-mail: ziweiz@princeton.edu.

# 1 Introduction

Principal component analysis (PCA), first introduced by Karl Pearson (Pearson, 1901), is one of the most commonly used techniques for dimension reduction in many disciplines, such as neurosciences, genomics and finance (Izenman, 2008). We refer the readers to Jolliffe (2014) for a recent review. It extracts latent principal factors that preserve most of the variation in the dataset. Let $\boldsymbol{x}$ be a random vector taking values in $\mathbb{R}^d$ with mean zero and covariance matrix $\boldsymbol{\Sigma}$. With this formalism, PCA seeks projection direction vectors, $\boldsymbol{v}_1, \ldots, \boldsymbol{v}_K \in \mathbb{R}^d$, such that

$$\boldsymbol{v}_1 \in \underset{\|\boldsymbol{v}\|_2=1}{\operatorname{argmax}} \boldsymbol{v}^\intercal \boldsymbol{\Sigma} \boldsymbol{v}, \ \boldsymbol{v}_2 \in \underset{\|\boldsymbol{v}\|_2=1, \boldsymbol{v} \perp \boldsymbol{v}_1}{\operatorname{argmax}} \boldsymbol{v}^\intercal \boldsymbol{\Sigma} \boldsymbol{v}, \ \boldsymbol{v}_3 \in \underset{\|\boldsymbol{v}\|_2=1, \boldsymbol{v} \perp \boldsymbol{v}_1, \boldsymbol{v}_2}{\operatorname{argmax}} \boldsymbol{v}^\intercal \boldsymbol{\Sigma} \boldsymbol{v}, \cdots.$$

In other words, $\{\boldsymbol{v}_j\}_{j=1}^K$ are the top $K$ eigenvectors of $\boldsymbol{\Sigma}$. Given $\mathbf{V}_K \equiv (\boldsymbol{v}_1, \ldots, \boldsymbol{v}_K)$, we can then project the original high dimensional data onto the low dimensional space spanned by columns of $\mathbf{V}_K$ to achieve the goal of dimensionality reduction. Since $\mathbf{V}_K$ captures the most variation in the dataset, these projected data points approximately preserve the geometric properties of the original data, which are amenable to downstream statistical analysis. In real applications, the true covariance matrix $\boldsymbol{\Sigma}$ is typically unknown; we need to substitute $\boldsymbol{\Sigma}$ with its empirical version $\widehat{\boldsymbol{\Sigma}}$.

The high complexity of big data, such as massiveness, contamination, nonlinearity and decentralization, has posed fundamental challenges to statistical inference. PCA, an 100-year-old idea, has been and is still shining as a powerful tool for modern data analytics. Modern developments for PCA focus on attempts that address various challenges created by big data. For example, the massiveness of features in big data has been shown to create many notorious problems, making many conventional inferential procedures ill-posed. Recent study (Fan, Liao and Mincheva, 2013) shows that PCA is closely connected to factor analysis. This motivates new methodological developments in multiple testing problems when tens of thousands of possibly dependent statistical tests are evaluated simultaneously (Fan, Han and Gu, 2012; Fan et al., 2017), which partially solve the high dimensional inference problem. Moreover, big data are often contaminated by outliers or heavy-tailed errors (Fan, Li and Wang, 2017; Sun, Zhou and Fan, 2017), motivating the use of robust covariance inputs in the PCA formulation (Fan, Wang and Zhu, 2016; Minsker, 2016). This results in a form of robust PCA. Moreover, machine learning algorithms, such as those in clustering, community detection, ranking, matrix completion and mixture models, often involve solving a highly nonconvex system. This makes developing practical and efficient computational algorithms a grand challenge. PCA, or spectral method more generally, can often be used to solve a reduction of the highly nonconvex system, without losing much statistical efficiency. See Anandkumar et al. (2014) and Abbe (2017). Manifold sometimes can be used to approximate the nonlinearity structure of a dataset. Surprisingly, PCA finds applications in this setting and achieves a form of nonlinear dimension reduction (Torgerson, 1952; Roweis and Saul, 2000).

# 2 Covariance matrix estimation and PCA

We now begin the journey of PCA for big data. Given a small number $K$, the main goal of PCA is to estimate the $K$-dimensional principal eigenspace of $\mathbf{\Sigma} \in \mathbb{R}^{d \times d}$ such that it captures most of the variation in the dataset. In statistical applications, the true covariance matrix $\mathbf{\Sigma}$ is typically unknown and in practice, PCA is conducted on some estimator $\widehat{\mathbf{\Sigma}} = \widehat{\mathbf{\Sigma}}(\mathbf{X}_1, \ldots, \mathbf{X}_n)$ of $\mathbf{\Sigma}$, where $\mathbf{X}_1, \ldots, \mathbf{X}_n$ are observations from $\mathbf{X}$. A significant error in recovering the eigenvalues and eigenvectors of $\mathbf{\Sigma}$ using those of $\widehat{\mathbf{\Sigma}}$ would lead to a substantial loss of information contained in the data by PCA projection. As direct applications of matrix perturbation theory (see Section 2.1), bounds on the estimation error $\|\widehat{\mathbf{\Sigma}} - \mathbf{\Sigma}\|$ under some matrix norm $\|\cdot\|$ can be effectively applied to relate the eigenvalues/eigenvectors of $\mathbf{\Sigma}$ and of $\widehat{\mathbf{\Sigma}}$ under the spectral gap condition. Therefore, it is important to build a good covariance estimator, say $\widehat{\mathbf{\Sigma}}$, with small statistical error in the sense that for any given $\delta > 0$, we are interested in minimizing the value $r$ that satisfies $\mathbb{P}(\|\widehat{\mathbf{\Sigma}} - \mathbf{\Sigma}\| > r) \leq \delta$ for some matrix norm $\|\cdot\|$.

## 2.1 Perturbation theory

Considering $\mathbf{\Delta} = \widehat{\mathbf{\Sigma}} - \mathbf{\Sigma}$ as a perturbation, it is crucial to understand how the eigenspace of $\widehat{\mathbf{\Sigma}} = \mathbf{\Sigma} + \mathbf{\Delta}$ perturbs around that of $\mathbf{\Sigma}$. This problem has been widely studied in the literature (Davis and Kahan, 1970; Wedin, 1972; Stewart, 1991, 2006; Yu, Wang and Samworth, 2015). Among these results, the $\sin \Theta$ theorems, established by Davis and Kahan (1970) and Wedin (1972), have become fundamental tools and are commonly used in numerical analysis, statistics and machine learning. While Davis and Kahan (1970) focused on eigenvectors of symmetric matrices, Wedin's $\sin \Theta$ theorem studies the singular vectors for asymmetric matrices and provides a uniform perturbation bound for both the left and right singular spaces in terms of the singular value gap and perturbation level. Over the years, various extensions have been made in different settings. For example, Vu (2011), Shabalin and Nobel (2013), O'Rourke, Vu and Wang (2013) and Wang (2015) considered the rotations of singular vectors after random perturbations; Cai and Zhang (2016) established separate perturbation bounds for the left and right singular subspaces. See also Dopico (2000) and Stewart (2006). Recently, Fan, Wang and Zhong (2016) derived new perturbation bounds, measured in the $\ell_\infty$-norm, for singular vectors (or eigenvectors in the symmetric case), which can be applied to robust covariance estimation in high-dimensional factor models and robust estimation of the false discovery proportion (Fan et al., 2017) when the sampling distributions are heavy-tailed. See Sections 3.2 and 3.3 for details on these statistical applications.

## 2.2 Robust covariance inputs and robust PCA

Due to matrix perturbation theory, upper bounds on the spectral norm $\|\widehat{\boldsymbol{\Sigma}} - \boldsymbol{\Sigma}\|_2$ or the elementwise $\ell_\infty$-norm (also known as the max norm) $\|\widehat{\boldsymbol{\Sigma}} - \boldsymbol{\Sigma}\|_{\max}$ can be used to establish corresponding bounds on the $\ell_2$ distance and $\ell_\infty$ distance between the population eigenvectors and their empirical counterparts obtained from $\widehat{\boldsymbol{\Sigma}}$, respectively. Given independent observations $\boldsymbol{X}_1, \ldots, \boldsymbol{X}_n$ from $\boldsymbol{X}$ with $\mathbb{E}\boldsymbol{X} = \boldsymbol{0}$, the sample covariance matrix, namely $\widehat{\boldsymbol{\Sigma}}_{\mathrm{sam}} := n^{-1} \sum_{i=1}^n \boldsymbol{X}_i \boldsymbol{X}_i^\mathsf{T}$, is arguably the most natural choice to estimate $\boldsymbol{\Sigma} \in \mathbb{R}^{d \times d}$ when the dimension $d$ is smaller than the sample size $n$. The finite sample bound on $\|\widehat{\boldsymbol{\Sigma}}_{\mathrm{sam}} - \boldsymbol{\Sigma}\|_2$ or $\|\widehat{\boldsymbol{\Sigma}}_{\mathrm{sam}} - \boldsymbol{\Sigma}\|_{\max}$ has been well studied in the literature (Vershynin, 2012; Tropp, 2012). If $\boldsymbol{X}$ has a sub-Gaussian distribution in the sense that for all unit vectors $\boldsymbol{v} \in \mathbb{S}^{d-1}$, $\mathbb{E}\exp(\lambda \boldsymbol{v}^\mathsf{T} \boldsymbol{X}) \leq \exp(c\lambda^2 \boldsymbol{v}^\mathsf{T} \boldsymbol{\Sigma} \boldsymbol{v})$ for some constant $c$, then Remark 5.40 in Vershynin (2012) implies that for every $t \geq 0$, $\|\widehat{\boldsymbol{\Sigma}}_{\mathrm{sam}} - \boldsymbol{\Sigma}\|_2 \leq \max(\delta, \delta^2) \|\boldsymbol{\Sigma}\|_2$ with probability at least $1 - 2e^{-t}$, where $\delta = C_1 \sqrt{(d+t)/n}$. Similarly, it can be shown that with probability greater than $1 - 2e^{-t}$, $\|\widehat{\boldsymbol{\Sigma}}_{\mathrm{sam}} - \boldsymbol{\Sigma}\|_{\max} \leq \max(\eta, \eta^2)$, where $\eta = C_2 \sqrt{(\log d + t)/n}$. Here, $C_1, C_2 > 0$ are constants depending only on $c$.

However, when the distribution is heavy-tailed, one cannot expect sub-Gaussian behaviors of the sample covariance in either the spectral or max norm (Catoni, 2012). Therefore, to perform PCA for heavy-tailed data, the sample covariance is a risky choice to begin with. Indeed, alternative robust estimators have been constructed to achieve better finite sample performance. In Fan, Li and Wang (2017), the authors constructed an elementwise robustified version of the sample covariance matrix using the adaptive Huber loss minimization (Sun, Zhou and Fan, 2017), and derived a sub-Gaussian-type deviation inequality in the max norm under finite fourth moment condition instead of sub-Gaussianity. Based on a novel shrinkage principle, Fan, Wang and Zhu (2016) and Minsker (2016) independently constructed global robustified variants of the sample covariance with sub-Gaussian behavior under the spectral norm as long as the fourth moment of $\boldsymbol{X}$ is finite. A different robust method using the idea of median-of-means was proposed and studied by Minsker (2015). More recently, Giulini (2017) studied robust PCA in a more general setting where the data sample is made of independent copies of some random variable ranging in a separable real Hilbert space. Together, these results provide a new perspective on robustness from a nonasymptotic standpoint, and also represent a useful complement to the previous results on robust PCA. For instance, Candés et al. (2011) focused on a different notion of robustness and showed that it is possible to recover the principal components of a data matrix when the observations are contained in a low-dimensional space but arbitrarily corrupted by additive noise. See also Chandrasekaran et al. (2011), Zhang and Lerman (2014) and the references therein.

# 3 PCA and factor analysis

Principal component analysis and factor analysis are two important problems in their respective fields and are seemingly unrelated at first sight. Lately, it is shown in Fan, Liao and Mincheva (2013) that the high-dimensional factor model is innately related to PCA, which makes it different from the classical factor model with fixed dimensionality (Lawley and Maxwell, 1971) and helps one understand why PCA can be used for the factor analysis when the top eigenvalues are spikes. In addition, this observation has triggered a series of interesting studies on matrix perturbation theory and robust covariance estimation.

## 3.1 Factor model and PCA

Let $\boldsymbol{X} = (X_1, \ldots, X_d)^\intercal$ be a random vector of outcomes of interest, which may represent financial returns, housing prices or gene expressions. The impact of dependence among outcomes is currently among the most discussed topics in various statistical problems, such as variable selection, covariance and precision matrix estimation and multiple testing. For example, financial returns depend on the equity market risks, housing prices depend on the economic health, gene expressions can be stimulated by cytokines, among others. Because of the presence of common factors, it is unrealistic to assume that many outcomes are uncorrelated. It is thus natural to assume a factor model structure, which relies on the identification of a linear space of random vectors capturing the dependence among the data. To do so, we consider an approximate factor model, which has been frequently used in economic and financial studies (Fama and French, 1992; Chamberlain and Rothschild, 1983; Bai and Ng, 2002) as well as genomics (Leek and Storey, 2008):

$$X_{ij} = \mu_j + \boldsymbol{b}_j^\intercal \boldsymbol{f}_i + \varepsilon_{ij}, \quad i = 1, \ldots, n, \, j = 1, \ldots, d, \qquad (3.1)$$

or in a matrix form $\boldsymbol{X}_i = \boldsymbol{\mu} + \mathbf{B} \boldsymbol{f}_i + \boldsymbol{\varepsilon}_i$ with $\mathbf{B} = (\boldsymbol{b}_1, \ldots, \boldsymbol{b}_d)^\intercal$.

Here $X_{ij}$ is the response for the $j$th feature of the $i$th observation $\boldsymbol{X}_i = (X_{i1}, \ldots, X_{id})^\intercal$, $\boldsymbol{\mu} = (\mu_1, \ldots, \mu_d)^\intercal$ is the intercept, $\boldsymbol{b}_j$ is a vector of factor loadings, $\boldsymbol{f}_i$ is a $K$-dimensional vector of common factors and $\boldsymbol{\varepsilon}_i = (\varepsilon_{i1}, \ldots, \varepsilon_{id})^\intercal$ is the error term, which is typically called the idiosyncratic component, uncorrelated with or independent of $\boldsymbol{f}_i$. We emphasize that, in model (3.1), only $X_{ij}$'s are observable. Intuitively, the unobserved common factors can only be inferred reliably when $d \to \infty$. Under model (3.1), $\boldsymbol{\Sigma} = \text{cov}(\boldsymbol{X})$ is given by

$$\boldsymbol{\Sigma} = \mathbf{B}\text{cov}(\boldsymbol{f}_i)\mathbf{B}^\intercal + \boldsymbol{\Sigma}_\varepsilon, \quad \boldsymbol{\Sigma}_\varepsilon = (\sigma_{\varepsilon,jk})_{1 \leq j,k \leq d} = \text{cov}(\boldsymbol{\varepsilon}_i). \qquad (3.2)$$

The literature on approximate factor models typically assumes that the first $K$ eigenvalues of $\mathbf{B}\text{cov}(\boldsymbol{f}_i)\mathbf{B}^\intercal$ diverge at rate $O(d)$, whereas all the eigenvalues of $\boldsymbol{\Sigma}_\varepsilon$ are bounded as $d \to \infty$. This assumption holds naturally when the factors are pervasive in the sense that a non-negligible fraction of factor loadings should be non-vanishing. The decomposition (3.2) is then asymptotically identifiable as $d \to \infty$.

Now we elucidate why PCA can be used for the factor analysis in the presence of spiked eigenvalues. Note that the linear space spanned by the first $K$ principal components of $\mathbf{B}\text{cov}(\boldsymbol{f}_i)\mathbf{B}^\intercal$ coincides with that spanned by the columns of $\mathbf{B}$ when $\text{cov}(\boldsymbol{f}_i)$ is non-degenerate. Therefore, we can assume without loss of generality that the columns of $\mathbf{B}$ are orthogonal and $\text{cov}(\boldsymbol{f}_i)$ is the identity matrix. See also Bai and Li (2012) and Bai and Ng (2013). Let $\overline{\boldsymbol{b}}_1, \ldots, \overline{\boldsymbol{b}}_K$ be the columns of $\mathbf{B}$, ordered such that $\|\overline{\boldsymbol{b}}_1\|_2 \geq \cdots \geq \|\overline{\boldsymbol{b}}_K\|_2$. Then, $\{\overline{\boldsymbol{\xi}}_j := \overline{\boldsymbol{b}}_j / \|\overline{\boldsymbol{b}}_j\|_2\}_{j=1}^K$ are the top $K$ eigenvectors of $\mathbf{B}\mathbf{B}^\intercal$ with eigenvalues $\{\overline{\lambda}_j := \|\overline{\boldsymbol{b}}_j\|_2\}_{j=1}^K$. Let $\{\lambda_j\}_{j=1}^d$ be the eigenvalues of $\boldsymbol{\Sigma}$ in a non-increasing order with the corresponding eigenvectors being $\{\boldsymbol{\xi}_j\}_{j=1}^p$. From the *pervasiveness* assumption that the eigenvalues of $d^{-1}\mathbf{B}^\intercal\mathbf{B}$ are distinct and bounded away from zero and infinity, it follows that $\{\overline{\lambda}_j\}_{j=1}^K$ grow at rate $O(d)$. In addition, by Weyl's theorem and the $\sin\Theta$ theorem of Davis and Kahan (Davis and Kahan, 1970),

$$\max_{1 \leq j \leq K} |\lambda_j - \overline{\lambda}_j| \bigvee \max_{j > K} |\lambda_j| \leq \|\boldsymbol{\Sigma}_\varepsilon\|_2 \quad \text{and} \quad \max_{1 \leq j \leq K} \|\boldsymbol{\xi}_j - \overline{\boldsymbol{\xi}}_j\|_2 = O(d^{-1}\|\boldsymbol{\Sigma}_\varepsilon\|_2).$$

These results state that PCA and factor analysis are approximately the same if $\|\boldsymbol{\Sigma}_\varepsilon\|_2 = o(d)$. This is assured through a sparsity condition on $\boldsymbol{\Sigma}_\varepsilon$, which is usually measured through $m_d = \max_{1 \leq j \leq d} \sum_{k=1}^d |\sigma_{\varepsilon,jk}|^q$ for some $q \in [0,1]$. The intuition is that, after subtracting out the common factors, many pairs of the cross-sectional units become weakly correlated. This generalized notion of sparsity was used by Bickel and Levina (2008), under which it holds that $\|\boldsymbol{\Sigma}_\varepsilon\|_2 \leq \max_{1 \leq j \leq d} \sum_{k=1}^d |\sigma_{\varepsilon,jk}|^q (\sigma_{\varepsilon,jj}\sigma_{\varepsilon,kk})^{(1-q)/2} = O(m_d)$ if the variances $\sigma_{\varepsilon,jj}$'s are uniformly bounded. Therefore, in the approximate sparse setting where $m_d = o(d)$, the pervasiveness assumption implies that the principal components $\{\lambda_j\}_{j=1}^K$ and the remaining components $\{\lambda_j\}_{j=K+1}^d$ are well separated and the first $K$ principal components are approximately the same as the standardized columns of the factor loading matrix. In this setting, PCA serves as a valid approximation to factor analysis only if $d \to \infty$.

## 3.2 Application to large covariance estimation

Covariance structure plays a particularly important role in high-dimensional data analysis. Apart from PCA, a large collection of fundamental statistical methods, such as linear and quadratic discriminant analysis, clustering analysis and regression analysis, require the knowledge of the covariance structure or certain aspects thereof. Realizing the importance of estimating large covariance matrices and the challenges that are brought by the high dimensionality, since the seminal work of Bickel and Levina (2008) and Fan, Fan and Lv (2008), various regularization techniques have been proposed to estimate $\boldsymbol{\Sigma}$ consistently. See Cai, Ren and Zhou (2016) and Fan, Liao and Liu (2016) for two comprehensive surveys on this topic. One commonly as-sumed low-dimensional structure is that the covariance matrix is sparse, namely many entries are zero or nearly so. In many applications, however, the sparsity assumption directly on $\boldsymbol{\Sigma}$ is not appropriate. A natural extension is conditional sparsity (Fan, Fan and Lv, 2008). Given the common factors, the outcomes are weakly correlated. In other words, in decomposition (3.2) we assume that $\boldsymbol{\Sigma}_\varepsilon$ is approximately sparse as in Bickel and Levina (2008). Motivated by the innate connection between PCA and factor analysis in high dimensions, Fan, Liao and Mincheva (2013) proposed a principal orthogonal complement thresholding method (POET) to estimate $\boldsymbol{\Sigma}$ in model (3.1): (i) Run the singular value decomposition on the sample covariance matrix of $\boldsymbol{X}$; (ii) keep the covariance matrix that is formed by the first $K$ principal components; and (iii) apply the adaptive thresholding procedure (Rothman, Levina and Zhu, 2009; Cai and Liu, 2011) to the remaining components of the sample covariance matrix.

As discussed in Section 2.2, in the presence of heavy-tailed data, the use of the sample covariance matrix is controversial and therefore is in question to begin with. Tailored to elliptical factor models, Fan, Liu and Wang (2017+) proposed to use the marginal Kendall's tau to estimate $\boldsymbol{\Sigma}$ and its top $K$ eigenvalues, and separately, use the spatial Kendall's tau to construct estimators for the corresponding leading eigenvectors. In more general settings where no shape constraints or parametric assumptions are imposed (normality, symmetry, elliptically symmetry, etc.), robust alternatives, such as the $U$-type covariance estimator and adaptive Huber covariance estimator considered in Fan et al. (2017), are preferred to be taken as the initial estimators in the POET procedure.

## 3.3 Application to factor-adjusted multiple testing

An iconic example of model (3.1) is the factor pricing model in financial economics, where $X_{ij}$ is the excess return of fund/asset $j$ at time $i$, $\boldsymbol{f}_i$'s are the systematic risk factors related to some specific linear pricing model, such as the capital asset pricing model (Sharpe, 1964), and the Fama-French three-factor model (Fama and French, 1993). Although the key implication from the multi-factor pricing theory is that the intercept $\mu_j$ should be zero, known as the "mean-variance efficiency" pricing, for any asset $j$, an important question is whether such a pricing theory can be validated by empirical data. In fact, a very small proportion of $\mu_j$'s might be nonzero according to the Berk and Green equilibrium (Berk and Green, 2004; Fan, Liao and Yao, 2015). It is practically important to identify those positive $\mu_j$'s, which amounts to conducting the following $d$ hypothesis testing problems simultaneously:

$$H_{0j} : \mu_j = 0 \quad \text{versus} \quad H_{1j} : \mu_j \neq 0, \quad j = 1, \ldots, d. \tag{3.3}$$

In the presence of common factors, $X_1, \ldots, X_p$ are highly correlated and therefore directly applying classical false discovery rate (FDR) controlling procedures (Benjamini and Hochberg, 1995; Storey, Taylor and Siegmund, 2004) can lead to inaccurate false discovery control and spurious outcomes. To improve the efficiency and to incorporate the strong dependence information, various factor-adjusted procedures have been proposed. See, for example, Barras, Scaillet and Wermer (2010), Fan, Han and Gu (2012) and Lan and Du (2017+). These works assume that both the factor and idiosyncratic noise follow multivariate normal distributions. However, the Gaussian assumption on the sampling distribution is often unrealistic in many applications, especially in genomics and finance. Also, as noted in Perry and Owen (2010), only non-Gaussian latent variables are detectable. To address the two challenges that are brought by strong dependence and heavy-tailedness, Fan et al. (2017) proposed a factor-adjusted robust multiple testing (FARM-Test) procedure with control of the false discovery proportion. Starting with a robust initial estimate of the covariance matrix, the FARM-Test method uses its first several principal components to estimate the factor loading matrix; next, using these estimated loading vectors, runs adaptive Huber regression to estimate the realized factors; robust test statistics are then constructed by subtracting out the realized common factors from the robust mean estimators; finally, the data-driven critical value is computed so that the estimated false discovery proportion is controlled at the prespecified level. The first step is motivated by the approximate equivalence between PCA and factor analysis as discussed in Section 3.2.

# 4 Applications to statistical machine learning

PCA finds applications in many statistical machine learning problems. In this section, we focus on its applications to four major problems in the machine learning literature: clustering and community detection, ranking, mixture model and manifold learning.

## 4.1 Clustering and Community Detection

Clustering and community detection is an important task in data analysis (Lowrimore and Manton, 2016). We focus on the stochastic block model (SBM), which is widely regarded as a canonical model for this purpose. We refer the readers to Abbe (2017) for a review of recent developments on SBM. We start with the definition of SBM. Let $n$ be the number of vertices, $k$ the number of communities, $\mathbf{p} = (p_1, ..., p_k)^\intercal$ the probability vector and $\mathbf{W}$ the $k \times k$ symmetric transition probability matrix with entries in $[0, 1]$. We say a pair $(\mathbf{X}, G)$ is sampled from $\text{SBM}(n, \mathbf{p}, \mathbf{W})$ if $\mathbf{X} = (X_1, \ldots, X_n)^\intercal$ is an $n$-dimensional random vector with independent and identically distributed components distributed under $\mathbf{p}$, and $G$ is an $n$-vertex simple graph where vertices $i$ and $j$ are connected with probability $W_{X_i, X_j}$, independently of other pairs of vertices. The $i$th community set is defined by $\Omega_i := \{v : X_v = i, v = 1, \ldots, n\}$. Roughly speaking, a SBM is a random graph with planted clusters: the cluster sizes follow a multinomial distribution with probability vector $\boldsymbol{p}$ and the probability that a member in the $i$ cluster and a member in the $j^{th}$ group get connected is $W_{i,j}$.

To fix idea, we consider the SBM with two communities, where the inner-cluster probability is $a$ and the across-cluster probability is $b$ with $b < a$. Further assume for simplicity that the two clusters have exactly size $n/2$, and index the first cluster with the first $n/2$ vertices. Given an observed graph with the adjacency matrix $\boldsymbol{A}$ (indicating whether or not two vertices are connected), the expected value $\mathbb{E}\mathbf{A}$ of this graph has four blocks given by

$$\mathbb{E}\mathbf{A} = \begin{bmatrix} a \cdot \mathbf{I}_{n/2} & b \cdot \mathbf{I}_{n/2} \\ b \cdot \mathbf{I}_{n/2} & a \cdot \mathbf{I}_{n/2} \end{bmatrix}.$$

It can be verified that the above matrix has three eigenvalues $a+b$, $a-b$ and $0$, where $0$ has multiplicity $n-2$, and the eigenvectors associated with the two largest eigenvectors are

$$\begin{bmatrix} \mathbf{1}_{n/2} \\ \mathbf{1}_{n/2} \end{bmatrix} \text{ and } \begin{bmatrix} \mathbf{1}_{n/2} \\ -\mathbf{1}_{n/2} \end{bmatrix},$$

respectively. In other words, if one were to work with the expected adjacency matrix, communities could be simply recovered by taking the eigenvector associated with the second largest eigenvalue. In practice, we only observe a single-shot of the SBM graph $\mathbf{A}$, which can be viewed as a perturbation of the expected adjacency matrix:

$$\mathbf{A} = \mathbb{E}\mathbf{A} + \mathbf{Z},$$

where $\mathbf{Z} = \mathbf{A} - \mathbb{E}\mathbf{A}$ is the perturbation. Therefore, the spectral method to achieve community detection can be formulated as solving the following optimization problem

$$\max_{\|\boldsymbol{x}\|_2^2 = n, \boldsymbol{x}^\intercal \mathbf{1}_n = 0} \boldsymbol{x}^\intercal \mathbf{A} \boldsymbol{x}, \text{ where } \mathbf{1}_n = (1,\ldots,1)^\intercal \in \mathbb{R}^n,$$

which is equivalent to finding the second eigenvector, as the first eigenvector is $\mathbf{1}_n/\sqrt{n}$.

However, as pointed by Abbe, Bandeira and Hall (2016), it has been an open problem whether the above spectral method is optimal for achieving the exact recovery property until very recently. We say that an algorithm achieves the exact recovery property if the community membership are recovered exactly with probability going to 1. Abbe et al. (2017) bridged this gap by providing a sharp entrywise eigenvector analysis for random matrices with low expected rank. Other than applications in SBM, the result can also be applied in other statistical machine learning problems, such as factor analysis and matrix completion, which we will not discuss in detail.

## 4.2 Ranking

Suppose we have a large collection of $n$ items and we are given partially revealed comparisons between pairs of items. For example, player A defeats player B; video A is preferred to video B when both are recommended. The goal is to identify the $K$ items that receive the highest ranks based on these pair comparisons. This problem, which is called the top-$K$ rank aggregation, has wide applications including web search (Dwork et al., 2001), recommendation systems (Baltrunas et al., 2010), sports competition (Massey, 1997), etc. One of the most widely used parametric models is the Bradley-Terry-Luce (BTL) model (Bradley and Terry, 1952). In the BTL model, we have

$$\mathbb{P}(\text{item } j \text{ is preferred over item } i) = \frac{\omega_j^*}{\omega_i^* + \omega_j^*},$$

where $\omega_i^*$ is the preference score of item $i$ or the ability of $i$-th person. The task then boils down to finding the $K$ items with the highest preference scores.

To see how PCA can be applied to this problem, we introduce one spectral method called "rank centrality" proposed by Negahban, Oh and Shah (2017). Consider each item as a node in a graph and construct a random walk on this graph where at each time, the random walk is possible to go from vertex $i$ to vertex $j$ if items $i$ and $j$ were ever compared; and if so, the likelihood of going from $i$ to $j$ depends on how often $i$ lost to $j$. That is, the random walk is more likely to move to a neighbor who has more "wins". The frequency this walk visits a particular node in the long run, or equivalently the stationary distribution, is the score of the corresponding item. Specifically, define the edge set $\mathcal{E} := \{(i,j) : \text{item } i \text{ and } j \text{ were compared}\}$ and consider the transition matrix $\mathbf{P}^* \in \mathbb{R}^{n \times n}$ such that

$$P_{i,j}^* = \begin{cases} \dfrac{1}{d} \dfrac{\omega_j^*}{\omega_i^* + \omega_j^*} & \text{if } (i,j) \in \mathcal{E}, \\ 1 - \dfrac{1}{d} \displaystyle\sum_{k:(i,k)\in\mathcal{E}} \dfrac{\omega_k^*}{\omega_i^* + \omega_k^*} & \text{if } i = j, \\ 0 & \text{otherwise}, \end{cases}$$

where $d$ is a sufficiently large constant that makes every row nonnegative (a proper probability distribution). We can verify that the normalized score vector

$$\boldsymbol{\pi}^* := \frac{1}{\sum_{i=1}^n \omega_i^*} [\omega_1^*, \ldots, \omega_n^*]^\mathsf{T}$$

is the stationary distribution of the Markov chain induced by $\mathbf{P}^*$, since $\mathbf{P}^*$ and $\boldsymbol{\pi}^*$ are in detailed balance, namely, $\pi_i^* P_{i,j}^* = \pi_j^* P_{j,i}^*$. Hence, by definition, $\boldsymbol{\pi}^*$ is the top left singular vector of $\mathbf{P}^*$: $\boldsymbol{\pi}^* \mathbf{P}^* = \boldsymbol{\pi}^*$. This motivates the algorithm of "rank centrality" that uses the top left singular vector of the empirical transition probability matrix $\widehat{\mathbf{P}}$ as an estimate of the preference score.

As for the statistical gaurantee, Negahban, Oh and Shah (2017) showed that $\Omega(n \log n)$ pairs are needed for consistency of estimating $\boldsymbol{\omega}^*$ in terms of the Euclidean norm. This is also the sample size needed for the Erdös-Rényi comparison graph to get connected, which is the minimum condition that makes the identification of top $K$ items possible. However, this does not lead to accurate identification of the top $K$ items. A recent work Chen et al. (2017) showed that the same sample complexity ensures exact top-$K$ identification, thus matching the minimax lower bound established by Chen and Suh (2015) before. This was accomplished via optimal control of the entrywise error of the score estimates.

## 4.3 Mixture Model

PCA can also be applied to learning latent variable models or mixture regression models, which are important models for investigating heterogeneous data. To illustrate the idea, we consider a mixture of $k$ Gaussian distributions with spherical covariances. Let $w_i \in (0,1)$ be the probability of choosing component $i \in \{1, \ldots, k\}$, $\{\boldsymbol{\mu}_1, \ldots, \boldsymbol{\mu}_k\} \subseteq \mathbb{R}^d$ be the component mean vectors, and $\sigma^2 \mathbf{I}$ be the common covariance matrix. An observation in this model is given by

$$\boldsymbol{x} = \boldsymbol{\mu}_h + \boldsymbol{z}, \tag{4.1}$$

where $h$ is a discrete random variable such that $\mathbb{P}(h = j) = w_j$ for $j = 1, \ldots, k$, and $\boldsymbol{z} \sim \mathcal{N}(0, \sigma^2 \mathbf{I})$. In other words, $\boldsymbol{x} \sim w_1 \mathcal{N}(\boldsymbol{\mu}_1, \sigma^2 \mathbf{I}) + \cdots + w_k \mathcal{N}(\boldsymbol{\mu}_k, \sigma^2 \mathbf{I})$ follows the mixture of the Gaussian distribution. We remark here that this is different from the topic model since every realization of $\boldsymbol{x}$ corresponds to a different realization of $h$. The parameters of interest are the component mean vectors, $\boldsymbol{\mu}_j$'s.

Let $\mathbf{M} = \mathbb{E}(\boldsymbol{x}\boldsymbol{x}^\intercal) - \sigma^2 \mathbf{I}$. Then, it is easy to see (Hsu and Kakade, 2013)

$$\mathbf{M} = \sum_{j=1}^{k} w_j \, \boldsymbol{\mu}_j \boldsymbol{\mu}_j^\intercal.$$

This indicates that running PCA on the population level would recover the $k$-dimensional linear subspace spanned by the mean component vectors. It can not fully identify all $\boldsymbol{\mu}_j$'s due to identifiability issues in general, but does help reduce the dimension of the parameter space, enabling the possibility of random initializations for a more delicate method in a second stage, such as higher-order tensor decomposition or likelihood based approaches. See Anandkumar et al. (2014) for a comprehensive review on tensor methods for latent variable models which includes the above discussed setting as a special case.

It is possible to extend the tensor decomposition method to study the finite mixture linear regression problems. One can replace the component mean vector in model (4.1) by $\langle \boldsymbol{x}, \boldsymbol{\beta}_h \rangle$, where $h$ is again a random variable indicating the label of submodels. We refer to Yi, Caramanis and Sanghavi (2016) and the references therein for more discussions on such generalizations. We point out that there are still many open problems in this direction. For example, the global landscape of mixture models are not well understood in general.

## 4.4 Manifold Learning

Some of the most popular methods in manifold learning are also based on spectral methods. For example, the classical multidimensional scaling (MDS) is used as a numerical tool for manifold learning, which is frequently used in psychometrics as a means of visualizing the level of similarity (or dissimilarity) of individual cases in a dataset (Torgerson, 1952). It is also known as principal coordinate analysis (PCoA), emphasizing the fact that the classical MDS takes the dissimilarity matrix as an input and outputs a coordinate matrix by assigning each object a coordinate. The classical MDS uses the fact that the coordinate matrix $\mathbf{U}$ can be derived by eigenvalue decomposition from $\mathbf{B} = \mathbf{U}\mathbf{U}^\intercal$, and the matrix $\mathbf{B}$ can be computed from proximity matrix $\mathbf{D}$ by using double centering:

1. Set up the matrix of squared dissimilarities $\mathbf{D}^2 = [d_{ij}^2]$.

2. Apply the double centering $\mathbf{B} = -\frac{1}{2}\mathbf{J}\mathbf{D}^2\mathbf{J}$ using the centering matrix $\mathbf{J} = \mathbf{I} - n^{-1}11^\intercal$, where $n$ is the number of objects.

3. Extract the $m$ largest positive eigenvalues $\lambda_1, \ldots, \lambda_m$ of $\mathbf{B}$ and the corresponding eigenvectors $\boldsymbol{v}_1, \ldots, \boldsymbol{v}_m$.

4. Let $\mathbf{U} = \mathbf{V}_m \boldsymbol{\Lambda}_m^{1/2}$, where $\mathbf{V}_m = (\boldsymbol{v}_1, \ldots \boldsymbol{v}_m)$ and $\boldsymbol{\Lambda}_m = \text{diag}(\lambda_1, \ldots, \lambda_m)$.

In the above algorithm, the dimension $m$ can be chosen using criteria based on the eigenvalue ratios such as those in the factor analysis (Ahn and Horenstein, 2013; Lam and Yao, 2012). The classical MDS is a linear dimension reduction method which uses the euclidean distances between objects as the dissimilarity measures. It has many extensions by designing different input matrices based on dataset and objectives. These extensions include the metric MDS, the nonmetric MDS and the generalized MDS, largely extending the scope of MDS, especially to the nonlinear setting. We refer readers to Borg and Groenen (2005) for various forms of MDS and corresponding algorithms. Theoretical analysis in the literature of MDS concentrates on the case whether the objects from a higher dimensional space can be embedded into a lower dimensional space, Euclidean or non-Euclidean. See, for example, de Leeuw and Heiser (1982). However, we are not aware of any statistical results measuring the performance of MDS under randomness, such as perturbation analysis when the objects are sampled from a probabilistic model.

Other nonlinear methods for manifold learning includes ISOMAP (Tenenbaum, de Silva and Langford, 2000), local linear embedding (LLE) (Roweis and Saul, 2000) and diffusion maps (DM) (Coifman and Lafon, 2006), to name a few. Most of these procedures relies on PCA or local PCA. We illustrate this using the LLE, which is designed to be simple and intuitive, and can be computed efficiently. It mainly contains two steps: the first step is to determine the nearest neighbors for each data point, and catch the local geometric structure of the dataset through finding the barycenter coordinate for those neighboring points; the second step is, by viewing the barycenter coordinates as the "weights" for the neighboring points, to evaluate the eigenvectors corresponding to the first several largest eigenvalues of the associated "affinity matrix" to locally embed the data to a lower dimensional Euclidean space. Surprisingly, despite its popularity in the manifold learning literature, Wu and Wu (2017) provided the asymptotic analysis of LLE until very recently.

## 5 Discussion

PCA is a powerful tool for data analytics (Jolliffe, 2014). Entering the era of big data, it is also finding many applications in modern machine learning problems. In this review article, we focus on clustering and community detection (Lowrimore and Manton, 2016), ranking, mixture model and manifold learning. Other applications, such as matrix completion (Keshavan, Montanari and Oh, 2010), phase synchronization (Zhong and Boumal, 2017), image segmentation (Dambreville, Rathi and Tannen, 2006) and functional data analysis (Ramsay, 2016), are not discussed here due to space limitations. Motivated by the fact that data are often collected and stored in distant places, many distributed algorithms for PCA have been proposed (Qu et al., 2002; Feldman, Schmidt and Sohler, 2013) and shown to provide strong statistical guarantees (Fan et al., 2017).

Another important application of PCA is augmented principal component regression, which is an extension of the classical principal regression (Kendall, 1957; Hotelling, 1957; Stock and Watson, 2002). The basic idea is to assume that latent factors that impact on a large fraction of covariates also impact on the response variable. Therefore, we use PCA to extract latent factors from the covariates and then regress the response variable on these latent factors along with any augmented variables, resulting in an augmented factor models. An example of this is the multi-index prediction based on estimated latent factors in Fan, Xue and Yao (2017). A related topic to this is the supervised principal component regression (Bair et al., 2006).

Recently, a number of interesting theoretical results on the empirical principal components under weak signals have been developed (Johnstone, 2001; Hoyle and Rattray, 2003; Baik, Ben Arous and Péché, 2005; Paul, 2007; Koltchinskii and Lounici, 2017; Wang and Fan, 2017), which are closely related to the rapid advances in random matrix theory. Interested readers are referred to the literature above for details.